\documentclass[aps,prx,10pt,notitlepage]{revtex4-1}

\usepackage{amsmath,amssymb,amsfonts,mathrsfs,graphicx}

\newcommand{\brho}{\bar{\rho}}
\newcommand{\cD}{{\cal D}}
\newcommand{\cH}{{\cal H}}
\newcommand{\cI}{{\cal I}}
\newcommand{\cZ}{{\cal Z}}
\newcommand{\cL}{{\cal L}}
\newcommand{\eg}{{e.g., }}
\newcommand{\ie}{{i.e., }}

\newcommand{\tpsi}{\tilde{\psi}}
\newcommand{\trho}{\tilde{\rho}}
\newcommand{\vecq}{{\bf q}}
\newcommand{\vecR}{{\bf R}}
\newcommand{\vecr}{{\bf r}}
\newcommand{\vecx}{{\bf x}}
\newcommand{\vecu}{{\bf u}}
\newcommand{\R}{{\mathbb R}}

\begin{document}

\title{Inferring entropy from structure}

\author{Gil Ariel}

\affiliation{Department of Mathematics, Bar-Ilan University, 52000
  Ramat Gan, Israel}

\author{Haim Diamant}

\affiliation{Raymond and Beverly Sackler School of Chemistry, Tel Aviv
  University, 69978 Tel Aviv, Israel}

\begin{abstract}

The thermodynamic definition of entropy can be extended to
nonequilibrium systems based on its relation to information. To apply
this definition in practice requires access to the physical system's
microstates, which may be prohibitively inefficient to sample or
difficult to obtain experimentally.  It is beneficial, therefore, to
relate the entropy to other integrated properties which are accessible
out of equilibrium. We focus on the structure factor, which describes
the spatial correlations of density fluctuations and can be directly
measured by scattering.  The information gained by a given structure
factor regarding an otherwise unknown system provides an upper bound
for the system's entropy. We find that the maximum-entropy model
corresponds to an equilibrium system with an effective
pair-interaction. Approximate closed-form relations for the effective
pair-potential and the resulting entropy in terms of the structure
factor are obtained. As examples, the relations are used to estimate
the entropy of an exactly solvable model and two simulated systems out
of equilibrium.  The focus is on low-dimensional examples, where our
method, as well as a recently proposed compression-based one, can be
tested against a rigorous direct-sampling technique. The entropy
inferred from the structure factor is found to be consistent with the
other methods, superior for larger system sizes, and accurate in
identifying global transitions. Our approach allows for extensions of
the theory to more complex systems and to higher-order correlations.

\end{abstract}

\maketitle


\section{Introduction}
\label{sec_intro}

After decades of slow progress, the theory of nonequilibrium
thermodynamics has seen substantial advances in recent years. One
outstanding example is the derivation of rigorous statistical
relations between macroscopic thermodynamic variables out of
equilibrium \cite{Jarzynski2011,Seifert2012}. Another is the concise
theoretical treatments of active matter \cite{Marchetti2013}.
The definition of thermodynamic state variables out of equilibrium
remains a key issue. Consistent definitions of temperature
\cite{CasasVazquez2003,Cugliandolo2011} and pressure \cite{Solon2015}
have proved problematic and context-dependent. By contrast, the
definition of entropy can readily be extended to nonequilibrium
systems based on its relation to
information \cite{Stratonovich1955,Parrondo2015}. Given the system's
set of microstates $\{s\}$ and their probability distribution $p_s$,
the physical Gibbs entropy is identified with the Shannon entropy,
\begin{subequations}
\begin{equation}
  H \equiv -\sum_s p_s \ln p_s,
\label{eq:entropy_disc}
\end{equation}
or its differential (continuous) analogue,
\begin{equation}
  H \equiv -\int p (s) \ln p(s) ds,
  \label{eq:entropy_cont}
\end{equation}
\label{eq:entropy}
\end{subequations}
where $p(s)$ is the probability density and the integration is over
its support.

As $H$ relates to (minus) the information encoded in $p_s$, it should
hold regardless of whether $p_s$ is an equilibrium distribution or
not \footnote{For a recent work that challenges this statement, see
S.\ Goldstein, J.\ L.\ Lebowitz, R.\ Tumulka, and N.\ Zanghi,
ArXiv:1903.1187.}. Equations \eqref{eq:entropy} are also used in a
range of fields and disciplines other than physics, including signal
and image processing \cite{Kerroum2010}, machine
learning \cite{Kwak2002,Zhu2010}, evaluation of
independence \cite{Calsaverini2009}, and independent component
analysis \cite{Faivishevsky2009}.

Beyond the definition of entropy, we need ways to compute or measure
it. In the case of physical systems this is useful, for example, for
identifying changes in the strength of correlations such as the ones
underlying global transitions. One may try to estimate $p_s$
sufficiently well to obtain $H$ directly from Eqs.~\eqref{eq:entropy}.
However, this essentially amounts to replacing a difficult problem
with a more difficult one \cite{Beirlant1997review}.  Moreover,
sampling $p_s$ also requires access to the physical system's
microstates, which may be hard or impossible to get experimentally.

Despite a large number of suggested algorithms, estimating the entropy
from independent sampling of the distributions remains a challenge,
especially for physical applications which typically involve
high-dimensional distributions in bounded
domains \cite{Beirlant1997review,Darbellay1999,Paninski2003}.  In such
cases, none of the commonly used direct estimation methods perform
well \cite{Beirlant1997review,Paninski2003,Ariel2020}.  It is
beneficial, therefore, to relate the entropy to other integrated
properties which are more directly accessible. At equilibrium such a
relation exists between the entropy (rather, its
temperature-derivative) and the fluctuations of the total energy
\cite{Book:Reif}. This relation assumes that $p_s$ is the Boltzmann
distribution, which is invalid out of equilibrium.

Density fluctuations are an integrated material property which is
directly accessible, both in and out of equilibrium, \eg via
scattering. Deviations from the density fluctuations of a completely
random particle distribution (an ideal gas) encode spatial
correlations and thus are related to the entropy associated with the
spatial distribution, the structural entropy. Our aim is to establish
this relation. For concreteness we restrict the discussion to
two-point density correlations in a system of identical particles at
steady state. The same approach, however, can be used to extend the
theory to more complex systems and to higher-order correlations. We
will discuss such extensions in Sec.~\ref{sec_discussion}. Denoting
the steady-state density of particles at position $\vecr$ by
$\rho(\vecr)$, the two-point correlation is
$\langle\rho(\vecr)\rho(\vecr')\rangle-\brho^2$, where
$\langle\cdot\rangle$ denotes an average over $p_s$ and $\brho
= \langle \rho(\vecr) \rangle$ is the mean density. Applying a Fourier
transform, we obtain the structure factor,
\begin{equation}
S(\vecq)=N^{-1}\langle |\trho(\vecq)|^2\rangle ,
\label{eq:S0}
\end{equation}
where $\tilde\rho(\vecq) = \int d\vecr e^{-i\vecq\cdot\vecr}
\rho(\vecr)$, and $N$ the number of particles.

Knowledge of the structure factor constrains the distribution of
microstates and thus entails an entropy cost. The relation between
entropy and structure factor is not unique; there may be many systems
having the same $S(\vecq)$ but different entropy. Still, we may ask
the sharply defined question, how much information is gained by
getting $S(\vecq)$ of an otherwise unknown system. Out of those many
systems sharing the same $S(\vecq)$, the answer is given by the one
which has the maximum entropy. Hence, for a given $S(\vecq)$, we can
get a bound for the system's entropy if we find the model which has
the largest entropy out of all possible $N$-body joint distributions
compatible with $S(\vecq)$. In Sec.~\ref{sec_maxH} we show that this
maximum-entropy model corresponds to an equilibrium system with an
effective pair-interaction dependent on the given $S(\vecq)$. Finding
such a pair-potential seems to be possible for any physically
realizable structure factor \cite{Zhang2020}. Additionally, a given
equilibrium structure factor can be associated with only one
pair-interaction \cite{Henderson1974}. Hence, the effective
pair-potential corresponding to the maximum-entropy system should
exist and is unique.  The problem is thus reduced to (a) finding the
effective pair-potential; and (b) calculating the resulting {\em
  equilibrium} entropy.

For step (b) we have the well-established arsenal of equilibrium
statistical mechanics. It is the first step, of determining the
effective pair-potential, which poses the extra challenge. In the
present work we prefer to remain on the level of analytic closed-form
expressions. Section~\ref{sec_eff} provides an approximate relation
for the entropy, which is our central result,
\begin{equation}
  h[S(\vecq)] = \frac{1}{2(2\pi)^d\brho} \int d\vecq \left( \ln S - S
  + 1 \right).
\label{eq:hS}
\end{equation}
Here, $h\equiv(H-H_{\rm id})/N$ is the excess entropy per particle
over the ideal-gas entropy, and $d$ the
spatial dimensionality.

Some of the points that we present appeared, from different
perspectives, in earlier studies. Other relations between the entropy
and spatial pair-correlations were derived in the past based on
equilibrium liquid theory
\cite{Green1947,Book:Green,Nettleton1958,Hernando1990,Laird1992}. We
will compare Eq.~\eqref{eq:hS} with those relations in subsequent
sections. The study in Ref.~\cite{Zhang2020} addressed the problem of
finding a system that realizes a given structure factor. The results
led to the conjecture that any realizable $S(\vecq)$ can be obtained
from an equilibrium system with an appropriate pair-interaction, and a
numerical scheme was developed to get the
pair-potential. Reference~\cite{Chakrabarty2011} studied universal
features of correlation functions in the limit of weak coupling and
how they could be used to extract microscopic pair-potentials from
measurements. We will point more specifically to the correspondence
between these earlier works and ours whenever relevant.

Recent works have proposed to overcome the sampling problem by
measuring the entropy of physical systems from their information
content, estimated through lossless data compression
\cite{AvineryPRL2019,MartinianiPRX2019,Melchert2015,EstevezRams2015}. We
will discuss this approach in detail later on.  Information content
was used also to identify elusive structures in amorphous materials
using network analysis \cite{Ronhovde2011}.

Entropy estimation based on the structure factor has several
advantages over the other available methods, in particular, when
applied to continuous physical systems. (a) Unlike sampling or data
compression, it does not involve any extrinsic discretization. Such
discretization is found to affect the results even when it is
performed in a generic parameter-free way \cite{Zu2020}. By contrast,
the Fourier modes $\vecq$ are naturally discretized by the system
size, which also allows for easy treatment of finite-size effects. In
addition, unlike the alternative relations mentioned above, which are
based on real-space pair correlations, Eq.~\eqref{eq:hS} does not
require binning of inter-particle distances. We find weak dependence
of Eq.~\eqref{eq:hS} on the upper cutoff of $\vecq$.  (b) Compression
algorithms, while being very simple to implement, depend on additional
extrinsic factors, such as transforming the data into a linear string,
and scaling of compression output to entropy \cite{Zu2020}. The former
issue may hamper the identification of long-range correlations,
whereas the structure factor encodes correlations of any range (albeit
on the pairwise level only).  See Ref.~\cite{Zu2020} for a recent
investigation into these issues.  Importantly, that study indicated
that the compression method might miss qualitative entropy changes of
certain driven systems which density correlations successfully picked
up.  (c) As to direct statistical estimation of entropy, the available
algorithms are inaccurate and prohibitively inefficient even at
moderate system sizes (100 particles and above). The
structure-factor-based estimation is not limited by system size.

A key disadvantage of entropy estimation based on density correlations
is that it is more particular. While sampling, and especially
compression, can be applied ``blindly'' without prior knowledge of the
system, relations such as Eq.~\eqref{eq:hS} are limited to systems of
a certain category. For example, Eq.~\eqref{eq:hS} must be modified if
it is to be applied to mixtures or to anisotropic particles.  In
addition, only approximate relations are available.  Yet, as
demonstrated below, Eq.~\eqref{eq:hS} is found to correctly indicate
qualitative features of strongly correlated as well as very small
systems.

We will compare the entropy inferred from the structure factor with
the results of two recently presented specific methods\,---\,one based
on direct sampling \cite{Ariel2020} and the other on data compression
\cite{AvineryPRL2019}. These methods will be briefly reviewed in
Sec.~\ref{sec_estimations}. The direct-sampling method is particularly
important for our purposes as it is guaranteed to yield accurate
results for small systems. This will allow us to test the
structure-factor and compression methods against a rigorous benchmark
out of equilibrium.

In Sec.~\ref{sec_examples} we apply all three methods to specific
systems. We have selected three examples. The first, treated in
Sec.~\ref{sec_gas}, is a simple equilibrium model of an ideal gas
partitioned into single-particle cells. In this system, whose entropy
is known exactly, density fluctuations are strongly suppressed due to
{\em single-body} potentials. Thus, it subjects our effective
pair-interaction approach to a stringent test. The second example
(Sec.~\ref{sec_Corte}) is a one-dimensional model inspired by
periodically sheared suspensions \cite{Corte2008}. This model
undergoes a transition to an absorbing state involving a sharp change
of entropy and strong density correlations. The third example,
described in Sec.~\ref{sec_RandOrg}, is a two-dimensional model of
random organization \cite{Hexner2015,Hexner2017}. With a
nonequilibrium transition involving a strongly correlated state, it
allows us to test the methods in two dimensions. We focus on small
system sizes (up to 100 particles), for which statistical estimation
of entropy is feasible.  Remarkably, the entropy reveals sharp
transitions between absorbing and non-absorbing steady-state dynamics
even for systems containing as little as 10 particles.  Finally, we
summarize our conclusions and discuss further applications and
extensions in Sec.~\ref{sec_discussion}.

\section{Entropy from structure factor}
\label{sec_entS}

\subsection{Maximum entropy model}
\label{sec_maxH}

In this section, we describe the maximum-entropy model for a given
structure factor, i.e., out of all models that have a given
$S(\vecq)$, we find the one that has the largest possible entropy. A
similar approach was used in Ref.~\cite{Zhang2020} to find a system
that realizes a given $S(\vecq)$. Here, we go further to derive an
expression for the maximum entropy.

We consider $N$ identical featureless particles in a volume $V$.  To
simplify notation, we assume a countable number of possible spatial
configurations $\vecR^s = (\vecR_1^s,\ldots,\vecR_N^s) \in \R^{Nd}$.
We denote the probability of configuration $\vecR^s$ by $p_s$. The
density field corresponding to configuration $\vecR^s$ is
$\rho^s(\vecr)=\sum_{n=1}^N \delta(\vecr-\vecR_n^s)$. Substitution in
Eq.~\eqref{eq:S0} gives the structure factor as
\begin{equation}
  S(\vecq)=N^{-1} \sum_s p_s \sum_{m=1}^N\sum_{n=1}^N
  e^{i\vecq\cdot(\vecR_m^s-\vecR_n^s)} .
\label{eq:appA:Sq_supp}
\end{equation}
Different sets of $p_s$ may have the same structure factor.  Given a
function $S(\vecq)$, we are looking for the specific model which has
the structure factor $S(\vecq)$ and a maximum value of $H=-\sum_s
p_s \ln p_s$. Note that $S(0)=N$ for any system; thus the mode $\vecq=0$
is excluded from the constrained $S(\vecq)$.

To find the maximum-entropy model, we rephrase the problem in terms of
Lagrange multipliers. We look for a maximum of $H$ under the infinite
set of constraints \eqref{eq:appA:Sq_supp} for all $\vecq \neq 0$, and
the normalization condition $\sum_s p_s=1$.  This general strategy is
not new; see, for example, Ref.~\cite{Jaynes1957}.  For simplicity of
notation, we use also a countable number of wavevectors $\vecq$.
We define the Lagrangian,
\begin{equation}
  \cL = H - \sum_{\vecq \neq 0} \lambda
     (\vecq) \left[ \sum_{m,n} \sum_s p_s
     e^{i\vecq\cdot(\vecR_m^s-\vecR_n^s)} - \sum_s p_s N
     S(\vecq) \right] - \mu \left[ \sum_s p_s - 1 \right] ,
\label{eq:appA:L}
\end{equation}
where $\lambda(\vecq)$ are the Lagrange multipliers associated with
each of the constraints \eqref{eq:appA:Sq_supp}, and $\mu$ is the
Lagrange multiplier associated with the normalization.  Then, the
maximum-entropy model satisfies,
\begin{equation}
   \frac{\partial \cL}{\partial p_s} = 0
   ~;~~~ \frac{\partial \cL}{\partial \lambda(\vecq)} = 0
   ~;~~~ \frac{\partial \cL}{\partial \mu} = 0 .
\end{equation}
Differentiation yields
\begin{subequations}
\begin{eqnarray}
    && -\ln p_s - 1 - \sum_{\vecq \neq 0} \lambda(\vecq)  \left[ \sum_{m,n} e^{i\vecq\cdot(\vecR_m^s-\vecR_n^s)} - N S(\vecq) \right]  - \mu =  0, \label{eq:appA:dp}\\
    &&  \sum_{m,n} \sum_s p_s e^{i\vecq\cdot(\vecR_m^s-\vecR_n^s)} = N S(\vecq),  \label{eq:appA:dlambda} \\
    && \sum_s p_s = 1. \label{eq:appA:dmu}
\end{eqnarray}
\label{eq:appA}
\end{subequations}
Solving Eq.~\eqref{eq:appA:dp} for $p_s$, we have
\begin{equation}
  p_s = e^{-(\mu+1)} \exp \left[ -  \sum_{\vecq \neq 0} \lambda(\vecq)  \left( \sum_{m,n} e^{i\vecq\cdot(\vecR_m^s-\vecR_n^s)} - N S(\vecq) \right)   \right]
  \equiv e^{-(\mu+1)} r_s ,
\label{eq:appA:pj}
\end{equation}
where $r_s$ is the exponential term.
Define,
\begin{equation}
  \cZ = \sum_s r_s = \sum_s \exp \left[ -  \sum_{\vecq \neq 0} \lambda(\vecq)  \left( \sum_{m,n} e^{i\vecq\cdot(\vecR_m^s-\vecR_n^s)} - N S(\vecq) \right)   \right] .
\label{eq:cZdisc}
\end{equation}
From Eq.~\eqref{eq:appA:dmu}, $\cZ=e^{\mu+1}$; hence,
\begin{equation}
   p_s = r_s / \cZ.
\label{eq:AppA:pjZ}
\end{equation}
In addition, 
\begin{eqnarray}
&&  \frac{\partial \ln \cZ}{\partial \lambda(\vecq)} = \cZ^{-1} \frac{\partial \cZ}{\partial \lambda(\vecq)} =
  -\cZ^{-1}  \sum_sr_s \left[ \sum_{m,n} e^{i\vecq\cdot(\vecR_m^s-\vecR_n^s)} - N S(\vecq) \right]  \nonumber \\
  && = -\sum_s p_s \left[ \sum_{m,n} e^{i\vecq\cdot(\vecR_m^s-\vecR_n^s)} - N S(\vecq) \right] = 
  -\sum_{m,n} \sum_s p_s e^{i\vecq\cdot(\vecR_m^s-\vecR_n^s)} + N S(\vecq) = 0 ,
\end{eqnarray}
where the last equality is due to Eqs.~\eqref{eq:appA:dlambda}
and \eqref{eq:appA:dmu}.  Thus, solving $\partial \ln \cZ
/ \partial \lambda(\vecq) = 0$ is equivalent to enforcing the
constraint on the structure factor.

Finally, we compute the entropy associated with the distribution
$p_s$,
\begin{eqnarray}
&&  H = -\sum_s p_s \ln p_s = - \sum_s p_s \ln \frac{r_s}{\cZ} = - \sum_s  p_s \ln r_s + \ln \cZ \nonumber \\
&& = -  \sum_s p_s \ln r_s + \ln \cZ .
\label{eq:appA:Hpj}
\end{eqnarray}
The first term satisfies,
\begin{eqnarray}
  && \sum_s p_s \ln r_s = \sum_s p_s \left[ -  \sum_{\vecq \neq 0} \lambda(\vecq)  \left( \sum_{m,n} e^{i\vecq\cdot(\vecR_m^s-\vecR_n^s)} - N S(\vecq) \right)   \right] \nonumber \\
  && = - \sum_{\vecq \neq 0} \lambda(\vecq) \sum_s p_s   \left[ \sum_{m,n} e^{i\vecq\cdot(\vecR_m^s-\vecR_n^s)} - N S(\vecq)  \right] \nonumber \\
  && =- \sum_{\vecq \neq 0} \lambda(\vecq) \left[  \sum_{m,n} \sum_s p_s e^{i\vecq\cdot(\vecR_m^s-\vecR_n^s)} - N S(\vecq)  \right] 
  .
\end{eqnarray}
The square brackets vanish for all $\vecq$.
We conclude that
\begin{equation}
  H = \ln \cZ .
\label{eq:appA:HlnZ}
\end{equation}

In summary, we see that the maximum-entropy model has the equilibrium
distribution of a system with an effective pair-potential
$\lambda(\vecq)$.  The probability of each
configuration \eqref{eq:appA:pj} is Boltzmann-like, and the entropy is
simply the log of the effective partition function $\cZ$.

\subsection{Effective pair-interaction and entropy}
\label{sec_eff}

We now use the results of Sec.~\ref{sec_maxH} to obtain the effective
pair-interaction and the resulting entropy estimation for a given
$S(\vecq)$. We will outline the calculation and give its results while
leaving the technical details to Appendix~\ref{appendix:Z}. Since we
are after the nominal entropy, we keep track of all prefactors.

Equation~(\ref{eq:cZdisc}) is rewritten in continuous form as
\begin{equation}
  \cZ = \int \left(\prod_{n=1}^N d\vecR_n\right)
  \exp \left[ -\int d\vecq \lambda(\vecq)
    \left( \sum_{m=1}^N\sum_{n=1}^N e^{i\vecq\cdot(\vecR_m-\vecR_n)} - NS(\vecq)
    \right) \right],
\label{ZR0}
\end{equation}
where we have absorbed the constant density of Fourier modes,
$\nu\equiv V/(2\pi)^d$, into the definition of the Lagrange
multipliers.  In Appendix~\ref{appendix:Z} we use standard methods to
transform Eq.~\eqref{ZR0} into the following functional integral over
a continuous field $\tpsi$,
\begin{eqnarray}
  \cZ &=& Z_{\rm id} (4\pi\nu^2)^{-\Omega_q/2} \exp
  \left[ \int d\vecq \left( N \lambda S -
    \frac{\nu}{2} \ln(\lambda/\nu) \right) \right]
  \int \cD\tpsi e^{-\cH_\psi},
\label{Zpsi0}\\
  \cH_\psi &=& -\brho \int d\vecR \left(
    e^{-i \int d\vecq \tpsi e^{-i\vecq\cdot\vecR}} - 1 \right) +\int d\vecq 
    \frac{|\tpsi|^2}{4\lambda}, \nonumber
\end{eqnarray}
where $Z_{\rm id}$ is the ideal-gas partition function, and $\Omega_q
= \nu\int d\vecq$ is the total number of Fourier modes.

Equation~\eqref{Zpsi0} is a useful starting point for finding the
effective interaction and the entropy to the desired order of
approximation. Here we confine ourselves to the lowest nontrivial
order which yields the simple relation, Eq.~\eqref{eq:hS}. This is
obtained by expanding $\cH_\psi$ to quadratic order in $\tpsi$ and
performing the Gaussian integration. The result is
\begin{equation}
  \cZ = Z_{\rm id} \exp \left[ \int d\vecq \left( N\lambda S -
    \frac{\nu}{2} \ln \left(1 + \lambda/\alpha \right) \right) \right],
\label{ZGaussian}
\end{equation}
where $\alpha\equiv\nu/(2N)=[2(2\pi)^d\brho]^{-1}$.

Applying the condition from Sec.~\ref{sec_maxH},
$\delta\ln\cZ/\delta\lambda(\vecq) = 0$, gives the effective
pair-interaction,
\begin{equation}
  \lambda(\vecq) = \alpha \left[ \frac{1}{S(\vecq)} - 1 \right].
\label{lambdaq}
\end{equation}
This relation between structure factor and pair-potential coincides
with earlier results obtained in different contexts
\cite{Chakrabarty2011,Zhang2020}. It is proportional also to (minus)
the direct correlation function known from equilibrium liquids via the
Ornstein-Zernike relation \cite{Book:Hansen}. Indeed, in the
high-temperature (i.e., weak-coupling) limit, the equilibrium direct
correlation function is proportional to (minus) the pair-potential
\cite{Book:Hansen}.  The effective potential~(\ref{lambdaq}) vanishes,
as expected, for an ideal gas ($S\equiv 1$). Substituting $\lambda$
back in $H=\ln\cZ$ leads to Eq.~\eqref{eq:hS}.  Inspection of
Eq.~\eqref{eq:hS} readily confirms that the excess entropy $h$ is
nonpositive for any $S(\vecq)$ and reaches its largest value of $0$
for $S(\vecq)\equiv 1$, \ie for an ideal gas.

\subsection{Comparison with previous expressions}
\label{sec_compare}

We now compare Eq.~\eqref{eq:hS} with previous results from
equilibrium liquid theory. At equilibrium, the excess entropy is
commonly extracted from simulations using the thermodynamic identity
\cite{Zu2020,Fomin2010},
\begin{equation}
  h = \frac{u}{T} - \int_0^{\brho} d\brho' \frac{p}{T\brho'^2},
\end{equation}
where $T$ is the temperature (taking the Boltzmann constant equal to
$1$), $u$ is the excess internal energy per particle over that of an
ideal gas, and $p$ the excess pressure over the ideal-gas
pressure. More relevant to the present work, however, are relations
between excess entropy and pair-correlations.

The first relation between the entropy and pair-correlations was
derived by Green \cite{Green1947,Book:Green},
\begin{equation}
  h^{\rm G} = -\frac{\brho}{2}
  \int d\vecr \left[ g\ln g - (g-1) \right],
\label{eq:hG}
\end{equation}
where $g(\vecr)$ is the pair distribution function, related to the
structure factor as
\begin{equation}
  S(\vecq) = 1 + \brho \int d\vecr e^{-i\vecq\cdot\vecr} [g(\vecr)-1].
\end{equation}
A higher-order relation was presented by Hernando
\cite{Hernando1990,Laird1992}\footnote{The expression derived in
  Ref.~\cite{Hernando1990} contains typos, which are corrected in
  Ref.~\cite{Laird1992}. Interestingly, this higher-order relation can
  be obtained directly from Green's 1947 article
  [Ref.~\cite{Green1947}, Eq.~(6.7)], when the pair distribution
  function is taken in its lowest-order approximation,
  $g(r)=e^{-\phi(r)/(k_{\rm B}T)}$, where $\phi(r)$ is the
  pair-potential.},
\begin{equation}
  h^{\rm H} = h^{\rm G} + \alpha \int d\vecq \left(
    \ln S - S + 1 + \frac{1}{2} (S-1)^2 \right).
\label{eq:hH}
\end{equation}
To compare Eq.~\eqref{eq:hS} with these expressions we expand
them around the ideal gas ($g=S=1$). For Green's functional,
\begin{equation}
  h^{\rm G} \simeq -\frac{\brho}{4} \int d\vecr (g-1)^2 + {\cal O}(g-1)^3
  = -\frac{1}{2}\alpha \int d\vecq (S-1)^2 + {\cal O}(g-1)^3.
\end{equation}
On the other hand, $\ln S - S + 1 = -\frac{1}{2}(S-1)^2 + {\cal
  O}(S-1)^3$. We conclude that the three expressions agree to second
  order in the correlations. In Secs.~\ref{sec_gas} and \ref{sec_RandOrg} they will be
  further compared for specific examples.

\section{Entropy estimations out of equilibrium}
\label{sec_estimations}

Despite a large number of suggested
algorithms \cite{Beirlant1997review,Paninski2003}, the problem of
estimating the entropy, in particular differential entropy, from
independent sampling of the distributions remains a challenge in high
dimensions \cite{Darbellay1999,Paninski2003}.  Broadly speaking,
traditional estimation methods can be classified into one of two
approaches\,---\,binning and sample-spacing methods, or their
multidimensional analogues\,---\,partitioning \cite{Stowell2009} and
nearest-neighbor methods \cite{Kozachenko1987,Lord2018}.  Traditional
partitioning methods work well at low dimensions (typically 2--3) and
only if the support is known.  In contrast, nearest-neighbor schemes
perform well up to moderately high dimensions (typically 10--15) and
for distributions with unbounded support.  However, they may fail
completely when the density has a compact support and their time
complexity grows exponentially with the dimension.

Overall, physical applications involving high-dimensional distributions
in bounded domains pose a difficult challenge.
Below, we review two recently proposed methods for estimating
differential Shannon entropy of compact high-dimensional continuous
distributions.

\subsection{Direct statistical estimation}
\label{sec_copula}

In this paper, we apply a recently suggested method that is based on
decomposing the distribution into a product of the marginal
distributions and the joint dependency, also known as the {\em
copula} \cite{Ariel2020}.  The entropy of the marginals is estimated
using one-dimensional methods, while the entropy of the copula, which
always has a compact support, is estimated recursively by splitting
the data along statistically dependent dimensions. Below we present
the method's principle, while the technical details of its application
are given in Appendix~\ref{appendix:entropy_estimation}. Further
information is found in Ref.~\cite{Ariel2020}.

Sklar's theorem \cite{Book:copula,Durante2010} states that any
continuous multi-dimensional probability density $p(\vecx)$,
$\vecx \in \R^D$, can be written uniquely as
\begin{equation}
       p(\vecx) = p_1(x_1) \cdot \dots \cdot p_D (x_D)
       c(F_1(x_1),\dots,F_D(x_D)) .
\label{eq:Sklar}
\end{equation}
Here, $\vecx=(x_1,\dots,x_D)$, and $p_k(\cdot)$ denotes the marginal
probability density of the $k$'th dimension, with cumulative
distribution function (CDF) $F_k (t) = \int_{-\infty}^t p_k(x) dx$.
The density of the copula, $c(u_1,\dots,u_D)$, is a probability
density on the hyper-square $[0,1]^D$, whose marginals are all uniform
on $[0,1]$,
\begin{equation}
       \forall k, ~\left[ \Pi_{j=1, ~j \neq k}^D \int du_j \right]
       c(u_1, \dots , u_D) = 1.
\label{eq:uniformMarginals}
\end{equation}
Substituting Eq.~\eqref{eq:Sklar} into the definition of
entropy, Eq.~\eqref{eq:entropy}, yields,
\begin{equation}
       H = \sum_{k=1}^D H_k + H_c,
\label{eq:Centropy}
\end{equation}
where $H_k$ is the entropy of the $k$'th marginal, to be computed
using appropriate 1D estimators, and $H_c$ is the entropy of the
copula, to be computed recursively by constructing a partition tree.

Splitting the overall estimation into marginal and copula
contributions has several advantages. First, the support of the copula
is compact, which is exactly the premise for which partitioning
methods are most adequate. Second, since the entropy of the copula is
nonpositive, adding up the entropy of marginals $H_k$ across
partition-tree levels provides an improving approximation (from above)
of the entropy.

\subsection{Compression-based estimation}
\label{sec_zip}

Following the ideas of Refs.~\cite{AvineryPRL2019,MartinianiPRX2019},
we estimate the entropy using a lossless compression algorithm.
Similar to Ref.~\cite{AvineryPRL2019}, samples are binned into 256
equal bins in each dimension, and the data are converted into a matrix
of 8-bit unsigned integers, in which each row constitutes an
independent sample.  The matrix is compressed using the LZW algorithm
(implemented in Matlab's imwrite function) into a Graphic Interchange
Format (gif) file. In order to estimate the entropy, the file size is
interpolated linearly between the file size of a compressed constant
matrix (minimum entropy) and a random matrix with independent
uniformly distributed values (maximum entropy), both of the same
dimensions.

The algorithm is extremely simple, which is its main appeal.  See
Appendix~\ref{appendix:entropy_estimation} for implementation details.

\section{Examples}
\label{sec_examples}

In this section we present the results of the three entropy estimation
methods, based on the structure factor, direct sampling, and
compression, for three simple models. The first is a 1D model of a
partitioned gas, in which the particles are restricted to
non-overlapping identical segments.  The second is a 1D model inspired
by periodically sheared particle suspensions \cite{Corte2008}. The
third is a 2D random organization (RandOrg)
model \cite{Hexner2015,Hexner2017}. The selected models all have
states of strongly suppressed density fluctuations
(hyperuniformity \cite{Torquato2018}) to put our density-correlation
approach to test \footnote{Note that, since in all examples $S(\vecq)$
vanishes as $\vecq \to 0$, the integral in Eq.~\eqref{eq:hS} is
singular (yet convergent).  Several numerical methods can integrate
over a singular point.  For simplicity, we use the trapezoidal rule
and simply set the integrand at $\vecq=0$ to zero.}. The partitioned
gas is an exactly solvable equilibrium model, while the other two
systems are out of equilibrium, checking the ability of the methods to
identify the entropy change due to a global dynamic
transition \cite{Zu2020}.

\subsection{Partitioned gas model}
\label{sec_gas}

We begin with a very simple model which is particularly suitable for
our purposes. It contains only single-particle constraints and thus is
exactly solvable. The exactly calculated entropy and structure factor
are far from those of an ideal gas due to strongly suppressed density
fluctuations (hyperuniformity). We compare our estimation methods
against these exact results. Since the entropy estimation based on
structure factor uses effective {\em pair}-interactions, this test is
particularly stringent.

\begin{figure}[h!tp]
\centering
\includegraphics[trim={1cm 9.5cm 1cm 1cm},clip, width=15cm, angle=0]{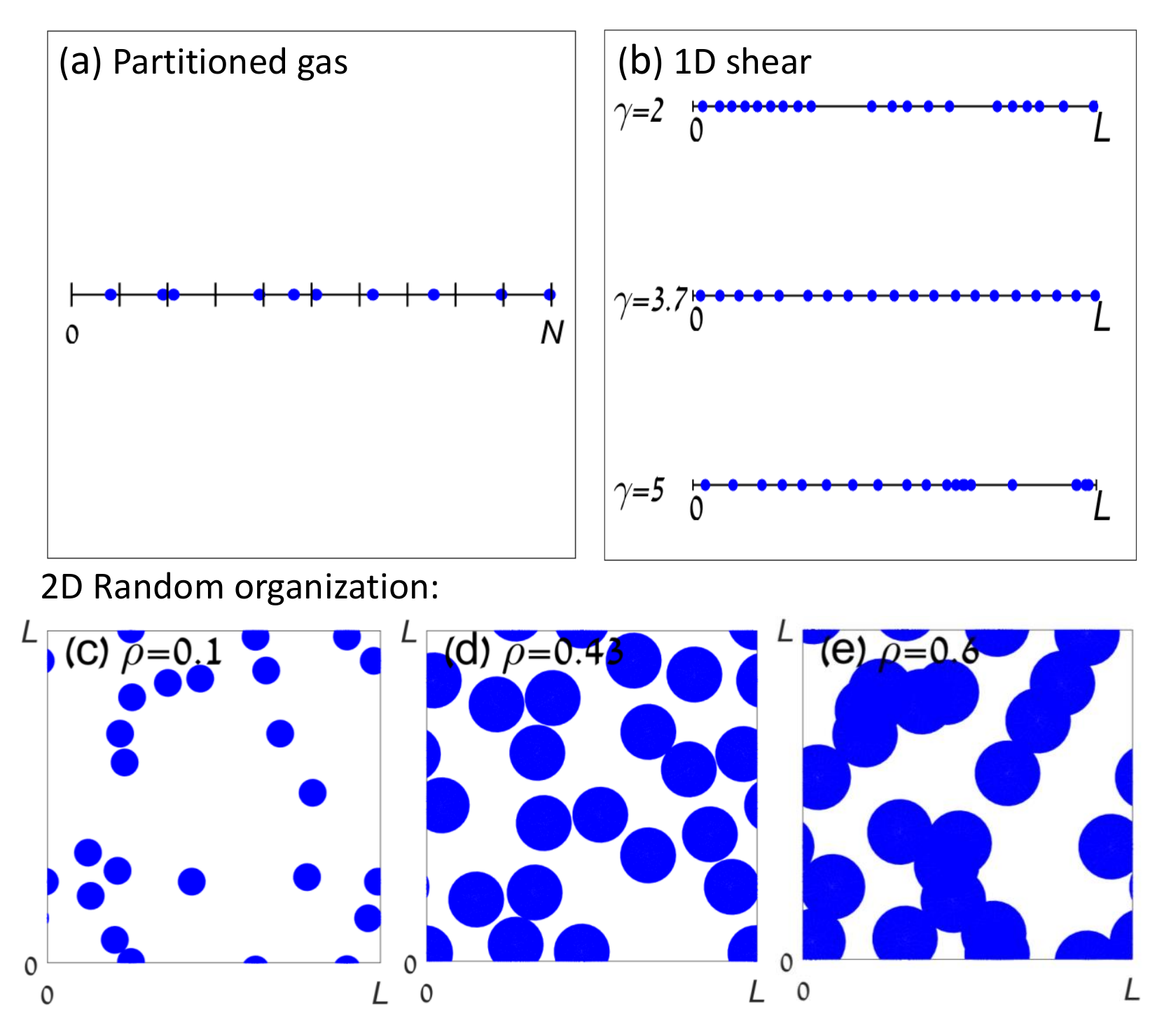}
\caption{
Snapshots from simulations.  {\bf a.} Partitioned gas model with
 $N=10$ particles. {\bf b.} The 1D shear model with $N=20$, below, at,
 and above the critical shear $\gamma_{\rm C}=3.7$.  {\bf c--e.} The
 2D RandOrg model with $N=20$ below, at, and above the critical density
 $\rho_{\rm C}=0.43$.  }
\label{fig:sketch}
\end{figure}

Consider $N$ point particles on a line of length $L$. The particles
are partitioned into $N$ cells of length $a=L/N$ each, one particle
per cell; see Fig.~\ref{fig:sketch}(a). Generalization to several
particles per cell is straightforward. We set $a\equiv 1$, such that
$L=N$, and the mean density is $\brho=1$. The entropy per particle is
\begin{equation}
  H_{\rm PG}/N = \ln a = 0,
\end{equation}
compared to the ideal entropy per particle, $\ln a+1=1$. Thus $h_{\rm
PG}=-1$.

To calculate the structure factor, we pick a random particle and place
the origin $x=0$ at the center of its cell. The particle's position,
$x_0\in(-1/2,1/2)$, divides the cell into two segments.
Denoting, $\ell_{\rm
min}=\min(x_0+1/2,1/2-x_0)$ and $\ell_{\rm
max}=\max(x_0+1/2,1/2-x_0)$, the probability per unit length to find a
second particle at the position $x$ is $0$ if $|x-x_0|<\ell_{\rm min}$,
$\brho$ if $|x-x_0|>\ell_{\rm max}$, and $\brho/2$ if
$|x-x_0|\in(\ell_{\rm min},\ell_{\rm max})$. Averaging over $x_0$,
we get the pair distribution function,
\begin{equation}
  g_{\rm PG}(x) = \left\{ \begin{array}{ll}
  |x|,\ \ \ \ & |x|\leq 1 \\
  1,\ \ \ \ & |x|>1.
  \end{array} \right.
  \label{eq:gPG}
\end{equation}
The structure factor is related to $g(x)$ as
$S(q) = 1 + \brho \int_{-\infty}^\infty dx e^{-iqx} [g(x)-1]$, yielding
\begin{equation}
  S_{\rm PG}(q) = 1 - \frac{4\sin^2(q/2)}{q^2}.
\label{eq:SPG}
\end{equation}
In the limit $q\ll 1$ we have $S_{\rm PG}\simeq q^2/12$. The decay to
zero for vanishing $q$ reflects
hyperuniformity \cite{Torquato2018}, \ie suppressed density
fluctuations over distances containing many cells.

Substituting Eq.~\eqref{eq:SPG} in Eq.~\eqref{lambdaq}, we get the (approximate)
effective pair-interaction,
\begin{equation}
  \lambda_{\rm PG}(q) = \frac{1}{\pi q^2} \, \frac{\sin^2(q/2)}
  {1 - 4\sin^2(q/2)/q^2}.
\label{lambdaPG}
\end{equation}
This is the fictitious Fourier-space pair-potential which mimics the
single-particle constraints. In the limit $q\ll 1$ it becomes
$\lambda_{\rm PG}(q)\simeq 3/(\pi q^2)$. Thus, over large distances,
the particles are distributed as in a 1D Coulomb gas.

Substituting $S_{\rm PG}(q)$ of Eq.~\eqref{eq:SPG} in
Eq.~\eqref{eq:hS} gives the entropy estimation $H/N=0.22$, compared to
the exact value of $0$ (or $h=-0.78$ compared to $-1$). Substitution
of $g_{\rm PG}(x)$ and $S_{\rm PG}(q)$ in the alternative relations
due to Green [Eq.~\eqref{eq:hG}] and Hernando [Eq.~\eqref{eq:hH}], gives
$H/N=0.75$ and $0.14$, respectively. These theoretical results
correspond to an infinite system, $N\rightarrow\infty$.

We simulated the partitioned-gas model numerically for different
numbers of particles, and applied the three entropy-estimation methods
to the resulting configurations.  Table~\ref{tbl:results10} presents
the results for $N=10$, $20$, $50$, and $100$ particles using $10^6$
independent samples.

As anticipated, the copula method performs well,
particularly at the lowest dimensions ($N=10$). Theoretically, both
the copula and compression methods must converge to the correct value
for an infinite number of samples (rightmost column of the table). In
practice, however, their accuracy is seen to deteriorate for larger
systems due to insufficient sampling of the enlarged phase space.

By contrast, the methods based on the structure factor and
pair-distribution function properly approach their individual
infinite-system limits given above. Larger system size implies finer
$q$-discretization, leading to better approximation of the structure
factor from the simulations. Note, on the other hand, that for the
partitioned-gas model these methods cannot give exact results, because
the model does not have actual pair-interactions. For this model we
find that our relation performs much better than Green's and slightly
worse than Hernando's. We recall that it has the advantage of using
$S(\vecq)$ alone, without the necessity of $g(\vecr)$, as will be
demonstrated further in Sec.~\ref{sec_RandOrg}.

\begin{table}[h]
\centering{}%
\begin{tabular}{|c|c|c|c|c|c|}
\hline
 & $N=10$ & $N=20$ & $N=50$ & $N=100$ & $N \to \infty$ \\
method  & $K=10^6$ & $K=10^6$ & $K=10^6$ & $K=10^6$ & $K \to \infty$  \\
\hline
copula (\ref{sec_copula}) & 4.8e-5 & 0.028 & $-0.0027$ & 0.093 & (0)\\
gif (\ref{sec_zip}) & 0.28 & 0.17 & $0.01$ & $-0.16$ & (0) \\
max entropy \eqref{eq:hS} & 0.45 & 0.39 & 0.31 & 0.28 & 0.22 \\
Green \eqref{eq:hG} & 0.54 & 0.63 & 0.69 & 0.72 & 0.75 \\
Hernando \eqref{eq:hH} & 0.31 & 0.28 & 0.22 & 0.19 & 0.14 \\
\hline
\end{tabular}
\caption{Estimations of entropy per particle for the partitioned-gas
model with $N$ particles and $K$ independent samples.  The exact value
for all systems is $H_{\rm PG}/N=0$ (compared to $H_{\rm id}/N=1$ for
an ideal gas).  The rightmost column (lower three rows) shows the
analytical estimations, valid for $N\rightarrow\infty$, using our
relation [Eq.~\eqref{eq:hS}] and two alternative ones
[Eqs.~\eqref{eq:hG} and \eqref{eq:hH}]. In the limit of infinite
sampling the copula and compression methods should become exact
(rightmost column, upper two rows). The rest of the entries in the
table contain results from numerical simulations.}
\label{tbl:results10}
\end{table}

As proved in Sec.~\ref{sec_maxH}, if the actual model is unknown and
only its $S(q)$ is given, then assuming an effective pair-potential is
imperative. Yet, on top of replacing the actual partitioning by a
fictitious pair-potential, our expression for that pair-potential in
terms of $S(q)$ [Eq.~\eqref{lambdaq}] is an approximation. We would
like to isolate and assess the effect of this approximation. To do so
we use a numerical optimization of the function $\lambda(q)$, similar
in spirit to the algorithm of Ref.~\cite{Zhang2020}. Simulated
particle configurations $\{R_n\}$ and trial values for $\lambda(q)$
are used to construct a distribution $p_s$ according to
Eq.~\eqref{eq:appA:pj}. This $p_s$ is used to obtain a structure
factor $S(q)$ according to Eq.~\eqref{eq:appA:Sq_supp}. Then, the
distance between the obtained $S(q)$ and $S_{\rm PG}(q)$ of
Eq.~(\ref{eq:SPG}) is minimized over the trial values of
$\lambda(q)$. Since any physical structure factor, including $S_{\rm
  PG}(q)$, can be tied to a unique pair-potential $\lambda(q)$, the
optimization in principle can get arbitrarily close to the exact
$\lambda(q)$, \ie to the exact maximum-entropy model. It is limited
only by the number of the discrete trial $\lambda$'s.  We note,
however, that the inverse problem of obtaining the pair-potential from
correlations is known to be notoriously difficult
\cite{Nguyen2017}. More details of the numerical procedure are found
in
Appendix~\ref{appendix:max_entropy_model}. Figure~\ref{fig:bestLambda}
shows the optimized $\lambda(q)$ for $N=10$ against the one found in
Eq.~\eqref{lambdaPG}.

\begin{figure}[h]
\centering
\includegraphics[trim={3cm 9cm 4cm 10cm},clip, width=10cm, angle=0]{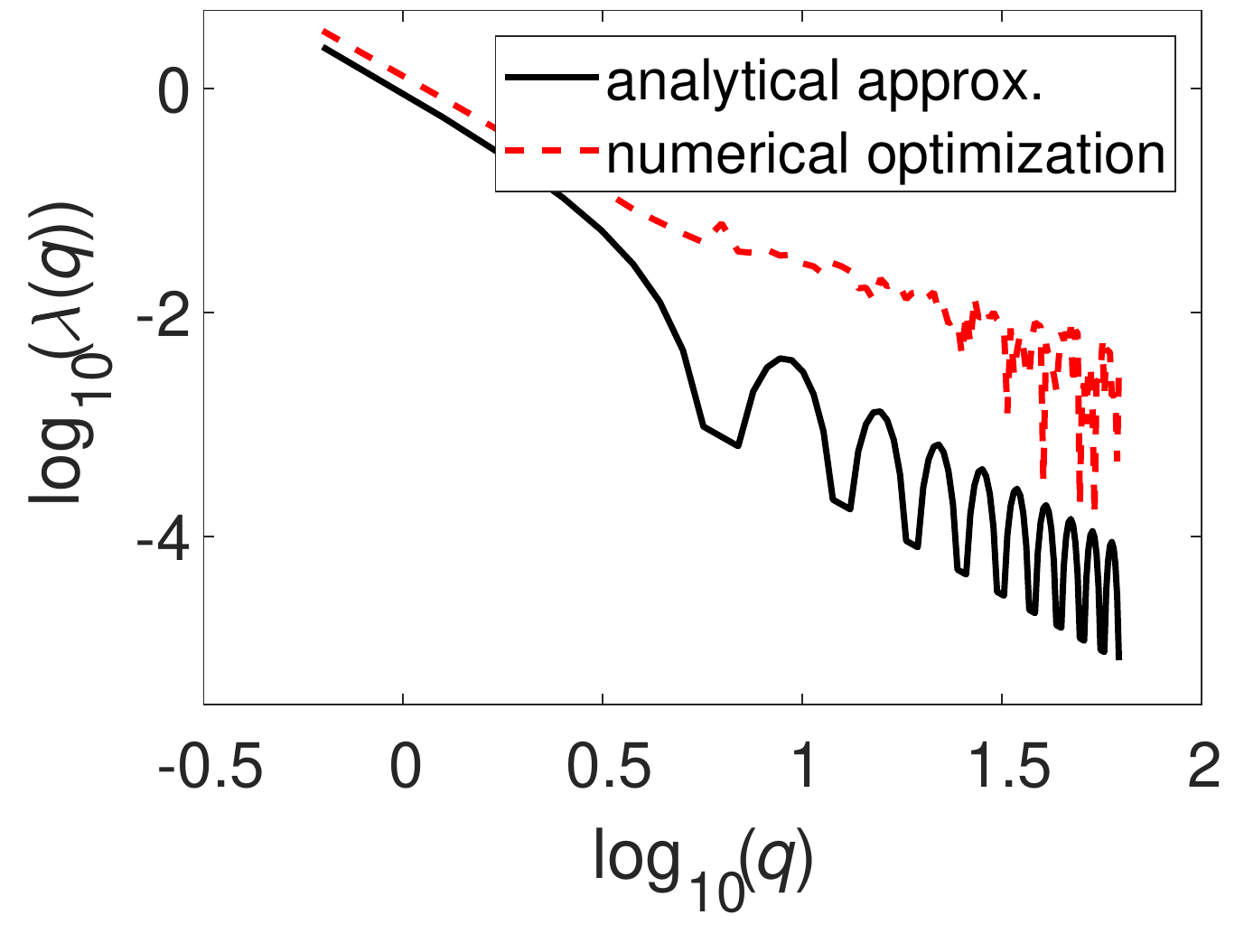}
\caption{Effective pair-potential for the partitioned-gas model. The lower black
curve shows the analytical approximate result
[Eq.~\eqref{lambdaPG}]. The upper red curve is the numerically
optimized pair-potential for $N=10$ particles, which yields a
structure factor as close as possible to the exact $S_{\rm PG}(q)$ of
Eq.~\eqref{eq:SPG}. See more details in
Appendix~\ref{appendix:max_entropy_model}.}
\label{fig:bestLambda}
\end{figure}

Substituting the numerically optimal $S(q)$ in Eq.~\eqref{eq:hS}, we get
the entropy of the maximum-entropy model while eliminating the
additional approximation leading to Eq.~\eqref{lambdaq}. For $N=10$ we
find $H/N=0.47$, to be compared with the value of $0.45$ obtained when
the structure factor from the simulations is used in Eq.~\eqref{eq:hS}
(see Table~\ref{tbl:results10}). Thus, despite the apparent
differences seen in Fig.~\ref{fig:bestLambda} between the approximate
pair-potential (\ref{lambdaPG}) and the numerically optimized one, the
ultimate effect on the entropy estimation is a mere $4.4\%$. We will
comment on this point in Sec.~\ref{sec_discussion}.

\subsection{One-dimensional shear model}
\label{sec_Corte}

We now examine a 1D model proposed in Ref.~\cite{Corte2008} to
schematically mimic periodically sheared suspensions. The model
consists of $N$ particles of diameter $\sigma$ placed at positions
$x_1,\dots,x_N \in [0,L]$. Boundary conditions are periodic.  At each
simulation time step, particles are displaced a distance
$\gamma \sigma$ to the right and back. Thus, a particle at position
$x_k$ will collide with other particles positioned within the segment
$[x_k-\sigma,x_k+\sigma(\gamma+1)]$.  At the end of each step,
colliding particles are randomly displaced by an independent normally
distributed shift with zero mean and variance $\epsilon \sigma$.  See
Fig.~\ref{fig:sketch}(b) for snapshots from simulations.

Varying $\gamma$ while keeping all other parameters ($N$, $\brho=N/L$,
$\sigma$ and $\epsilon$) fixed, one finds two distinct regimes. For
$\gamma$ smaller than a critical threshold $\gamma_C$, the system
reaches an absorbing state in which no collisions occur.  The time
required to reach the final state diverges as $\gamma \to \gamma_C$.
For $\gamma>\gamma_C$, the average rate of collisions per step does
not vanish but tends to a 
constant.  Interestingly, the critical threshold $\gamma_C$ is smaller
than the maximal density for which there exists a configuration with no collisions, 
$(\gamma+2)\sigma/L$.  
This implies that the system does not reach an absorbing state even though
there are configurations with no overlaps. For $\gamma=\gamma_C$, the
distribution of particle positions is hyperuniform, \ie strongly
correlated, with a structure factor that vanishes as $q\to
0$ \cite{Corte2008,Hexner2015,Hexner2017}.

To estimate the entropy of the 1D shear model, $10^6$ realizations
with independent initial conditions were run with parameters
$\sigma=1$, $\epsilon=1/2$, $\brho=0.2$, $N=10$, $20$, $50$, $100$,
and $\gamma$ in the range $[0,6]$.  Steps were repeated until an
absorbing state was reached (no collisions) or the collision rate
changed by less than 1\% in the second half of the simulation.  Thus,
for every $N$ and $\gamma$, we obtain $10^6$ independent samples from
the steady-state $D=N$-dimensional distribution of $(x_1,\dots,x_N)$.
The samples are used to estimate the entropy by applying the three
methods described above.  To decrease the phase-space volume, all
samples were sorted in increasing order along the $x$-axis prior to
estimation. We also show results for $N=1000$ using $10^5$
realizations.

\begin{figure}[h!tp]
\centering
\includegraphics[trim={0cm 3.5cm 0cm 0cm},clip, width=15cm, angle=0]{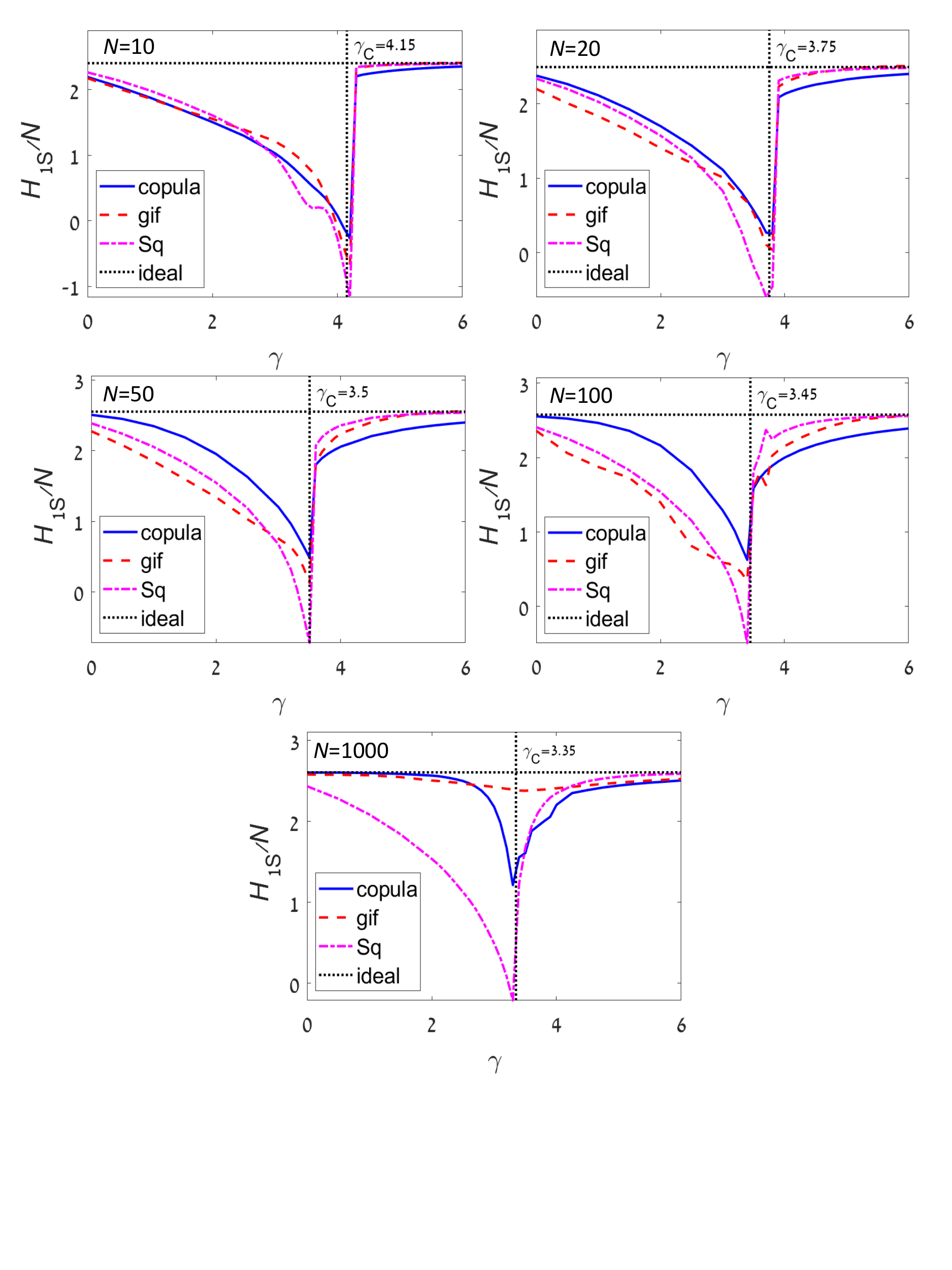}
\caption{ 1D shear model: Entropy per particle with $N=10$, $20$,
  $50$, $100$, and $1000$ particles. For the smaller systems all
  methods show a sharp change in entropy at the critical shear
  $\gamma_C$ as inferred from the largest value of $\gamma$ for which
  an absorbing state is reached (vertical dashed line).  The
  horizontal dashed line shows the ideal-gas entropy.  }
\label{fig:Corte_results}
\end{figure}

Figure~\ref{fig:Corte_results} shows the entropy estimations. For
$N\leq 100$ all three methods yield a sharp change in entropy at the
critical shear $\gamma_C$, as obtained independently by marking the
largest value of $\gamma$ for which an absorbing state is
reached. Note that, due to the small system sizes, $\gamma_C$ varies
with $N$. The horizontal dashed line shows the ideal-gas entropy.  At
higher phase-space dimensions (larger $N$), estimations from the
copula and compression methods become noisy and fluctuating, whereas
the results of the structure-factor method become smoother. For
$N=1000$ the compression method no longer captures the
transition. Overall, since the exact entropy is unknown, we cannot
definitely say which method is more accurate. For small $N$, however,
we expect the direct sampling (copula method) to give the most
reliable results, as confirmed in Sec.~\ref{sec_gas}.

Thus, the entropy estimations using the structure factor and
compression are in line with direct sampling. Our results demonstrate
that entropy estimation is a robust method for detecting global
transitions out of equilibrium even for small systems. 
Surprisingly, the sharp change in entropy is
evident even for very small systems, containing just 10 particles.

\subsection{Two-dimensional random organization model}
\label{sec_RandOrg}

Next, we check a 2D random organization (RandOrg) model. The model
consists of $N$ particles of diameter $\sigma$, placed in a
rectangular domain $[0,L]^2$ with periodic boundaries.  We implement a
zero-shear version of the model \cite{Hexner2015,Hexner2017}, in which
overlapping particles, termed `active', are displaced randomly.  The
displacements are uniformly distributed in a small circle with radius
$\epsilon$. When varying the area fraction $\rho=\pi \sigma^2 /4L^2$ while
keeping the rest of the parameters fixed, two regimes are observed,
similar to the 1D shear model.

For $\rho$ smaller than a critical threshold $\rho_C$, the system
reaches an absorbing state in which no particles overlap.  The time
required to reach the final state diverges as $\rho \to \rho_C$.  For
$\rho>\rho_C$, the average rate of random displacements per step does
not vanish but tends to a 
constant.  Again, $\rho_C$ is smaller than the maximum packing ratio
for circles ($\sqrt{3} \pi/6 \simeq 0.907$).  For $\rho=\rho_C$, the
distribution of particle positions is hyperuniform, with a structure
factor that vanishes as $q \to 0$ \cite{Hexner2015,Hexner2017}.  See
Fig.~\ref{fig:sketch}(c)--(e) for snapshots from simulations, below,
at, and above the critical area fraction.

In order to estimate entropy in the 2D RandOrg model, $10^6$
realizations with independent initial conditions were run with
parameters $\sigma=1$, $\epsilon=1/2$, $N=10, 20, 50$, and $\rho$ in
the range $[0.01,0.6]$.  Steps were repeated until an absorbing state
was reached (no collisions) or the collision rate changed by less than
1\% in the second half of the simulation.  Thus, for every $N$ and
$\rho$, we obtain $10^6$ independent samples of the $D=2N$-dimensional
distribution of $(x_1,\dots,x_N,y_1,\dots,y_N)$.  The samples are used
to estimate the entropy by applying the three methods described above.
Again, in order to decrease the phase space volume, all samples were
sorted by their $x$ component in increasing order prior to estimation.
The order of the $x$ and $y$ coordinates,
e.g. $(x_1,\dots,x_N,y_1,\dots,y_N)$ or $(x_1,y_1,\dots,x_N,y_N)$, did
not change the results.

\begin{figure}[h!tp]
\centering
\includegraphics[trim={0cm 0cm 0cm 0cm},clip, width=16cm, angle=0]{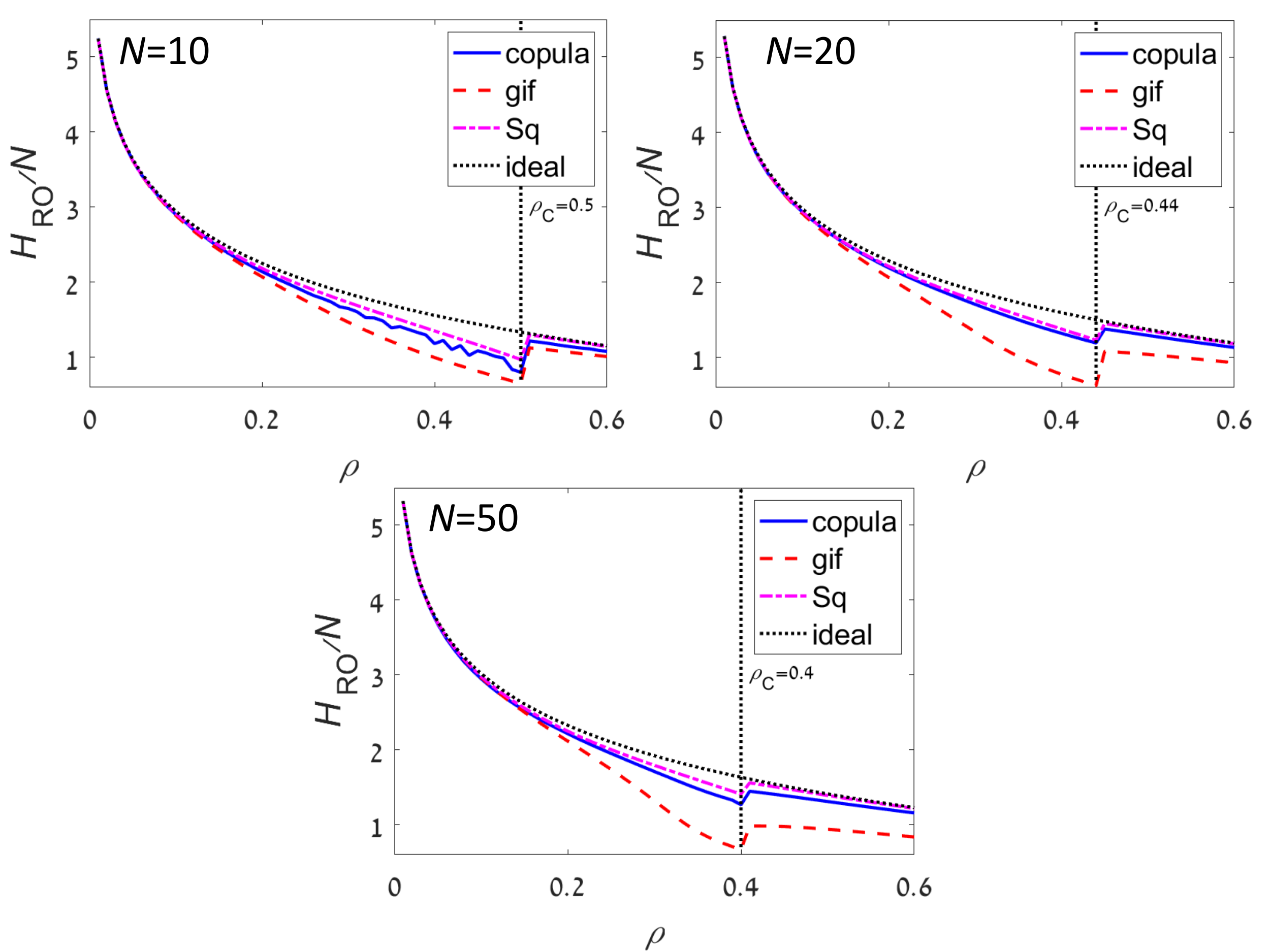}
\caption{
2D RandOrg model: Entropy per particle for $N=10$, $20$, and $50$
particles (corresponding to a multidimensional phase space of $20$,
$40$, and $100$ dimensions, respectively).  All methods show a small
sharp change in entropy at the critical density $\rho_C$ (vertical
dashed line) inferred from the largest value of $\rho$ for which an
absorbing state is reached. The dashed curve shows the ideal-gas
entropy.  }
\label{fig:RandOrg_results}
\end{figure}

Figure~\ref{fig:RandOrg_results} shows the entropy estimations for the
2D RandOrg model.  Similar to the 1D shear model, all methods yield a
a sharp change in entropy at the critical density $\rho_C$; yet, the
change is smaller than in the 1D case. Here, the structure-factor
estimations lie closer to the copula results than the
compression-based ones (which are lower).  Note that the jumps seen in
Fig.~\ref{fig:RandOrg_results} are significantly larger than those
reported in Ref.~\cite{MartinianiPRX2019} for a similar model (see
Fig.~4 there). In Ref.~\cite{MartinianiPRX2019} the entropy was
estimated using the lossless compression approach. The method was
applied, however, to a single sample of a significantly larger system
[about $(2$--$20)\times 10^3$ particles). This difference demonstrates
  the advantage of considering systems with a relatively small number
  of particles when evaluating entropy based on data compression.

\begin{figure}[h!tp]
\centering
\includegraphics[trim={0cm 11cm 0cm 0.5cm},clip, width=16cm, angle=0]{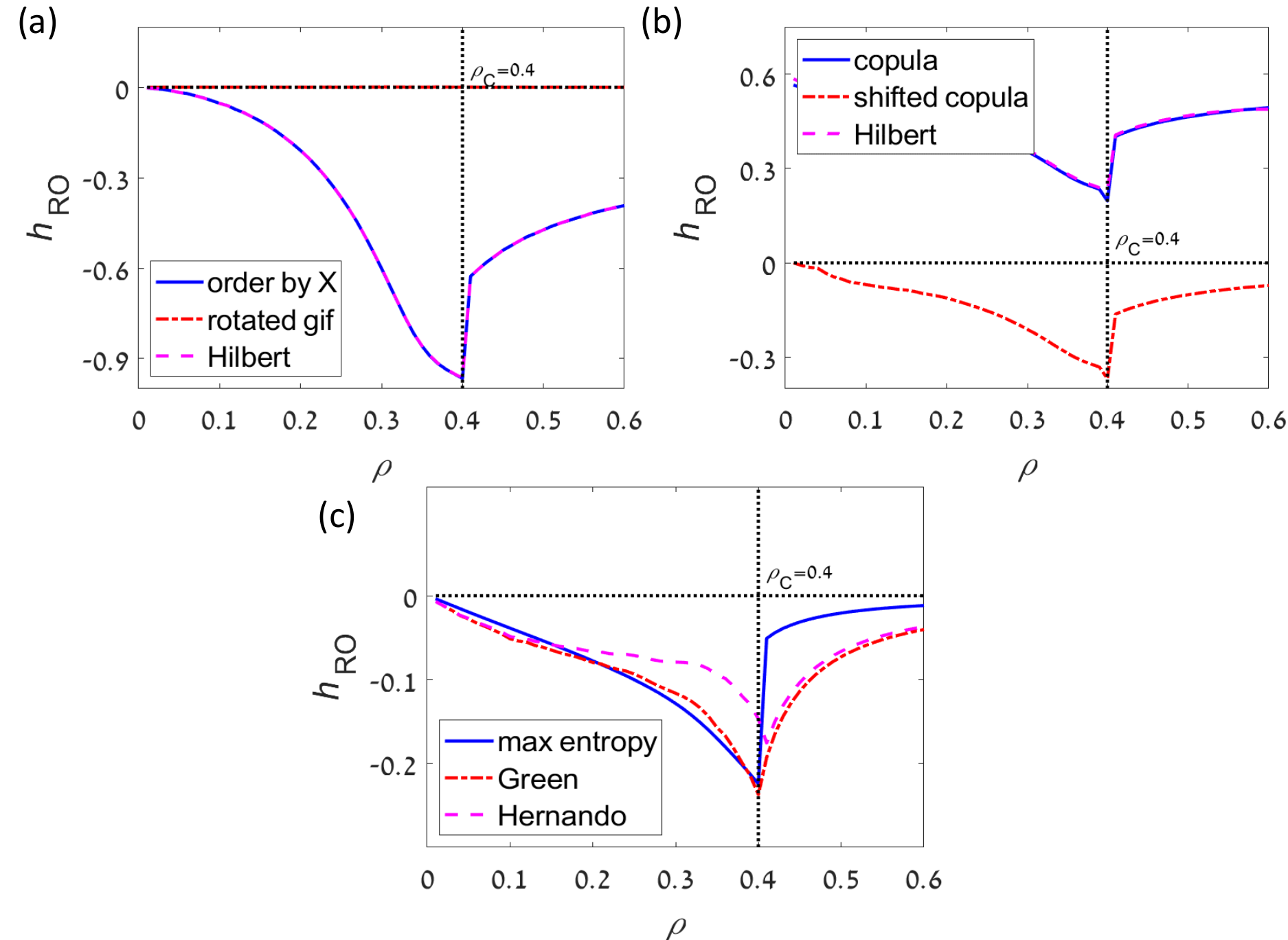}
\caption{
Method issues. (a) Copula method: the raw results are shifted to match
the ideal-gas limit.  (b) Compression method: the results are highly
sensitive to the order of independent samples in the compressed data
matrix (red line), but not to the order of points within each sample
(magenta line showing reordering by a Hilbert scan).  (c) Structure
factor: comparing different entropy--structure relations. All panels
present the excess entropy per particle for the 2D RandOrg model with
$N=50$.  }
\label{fig:discuss_RandOrg}
\end{figure}

We further use the 2D RandOrg results, for which statistical
estimation of the entropy is more challenging than in 1D, to
demonstrate some of the technical issues involved in the estimations.
We begin with the copula method.  Both the compression and
structure-factor methods use the ideal-gas case as a reference value;
they are, by definition, exact for an ideal gas.  In contrast, direct
estimation methods such as copula splitting are general-purpose
estimators and do not make such assumptions.  Hence, we treat the
estimation of the copula method for the ideal-gas case as a systematic
error, subtracting it from all the results.  The difference between
the raw results and the shifted ones is depicted in
Fig.~\ref{fig:discuss_RandOrg}(a).
Note that this error significantly increases with the dimension, growing from practically negligible at $N=10$, to the values depicted for $N=50$ in Fig.~\ref{fig:discuss_RandOrg}(a).

In the compression-based method, as explained in Sec.~\ref{sec_zip},
the data are converted into a matrix of 8-bit unsigned integers, in
which each row constitutes an independent sample.  It is well known
that the order of the compressed data can have a significant effect on
the compression ratio.  For example, it is more efficient to compress
text files by scanning columns rather than rows, because columns tend
to be more repetitive.  Similarly, in our application, transposing the
data matrix yields significantly worse compression ratios because rows
are independent, which implies that correlations among in-sample data
are long-ranged. Figure~\ref{fig:discuss_RandOrg}(b) shows entropy
estimates for the 2D RandOrg model obtained by transposing the data
matrix. This small change causes the method to perform poorly, and the
entropy at all surface fractions is practically indistinguishable from
that of an ideal gas.  We also tested ordering the samples along a
Hilbert scan of the 2D square domain \cite{MartinianiPRX2019}.  Our
numerical examples show no advantage of the Hilbert scan compared to
simple sorting along one dimension (see
Figs.~\ref{fig:discuss_RandOrg}(a) and \ref{fig:discuss_RandOrg}(b)).

Finally, Fig.~\ref{fig:discuss_RandOrg}(c) compares our relation,
Eq.~\eqref{eq:hS}, with the alternative relations between entropy and
pair-correlations, Eqs.~\eqref{eq:hG} and \eqref{eq:hH}. We see that
all expressions agree qualitatively (recall that Green's expression
performed poorly in the partitioned-gas example). The expression due
to Hernando slightly misses the location of the critical
density. Equation~\eqref{eq:hS} produces smoother and sharper results,
partly due to the binning of inter-particle distances required for the
other two relations.

\section{Discussion}
\label{sec_discussion}

One of this work's aims has been to put the measurement and
computation of entropy out of equilibrium on firmer ground. We have
proved that spatial density correlations, captured by the directly
measurable structure factor, can be used to obtain a rigorous upper
bound for the entropy of a particulate system. To infer the upper
bound from structure, we have used a new approximate relation
[Eq.~\eqref{eq:hS}] which, while giving comparable results to known
relations, has the advantage of depending only on the Fourier-space
structure factor [cf.\ Fig.~\ref{fig:discuss_RandOrg}(c)]. We have
tested the structure-based entropy estimation, together with two
additional methods based on either direct sampling or data
compression, on three systems in and out of equilibrium. The
direct-sampling (copula) method has offered a reliable benchmark for
finding the entropy of small systems out of equilibrium (a feature
that was missing in earlier studies).

Overall, the three methods have yielded qualitatively consistent
estimations, and successfully identified nonequilibrium transitions.
The statistical sampling and compression methods become less accurate
for larger systems, because sampling high-dimensional distributions
requires excessively large number of samples.  The computational
complexity of statistical sampling methods, such as the copula
splitting applied here, severely limits their applicability even for
moderate system sizes (up to 100 particles). Compression methods can
be applied to much larger systems; yet, they do not bypass the
sampling issue.  In comparison, the structure-based method does not
have this limitation. In the considered examples, for small systems,
the results of the structure-based method are closer to the
direct-sampling estimation, which may indicate better accuracy
compared to data compression. We conclude that consistent entropy
estimation out of equilibrium is achievable, with significant
advantages to doing so based on the structure factor whenever this
method is relevant.

Two major approximations underlie Eq.~\eqref{eq:hS}. First, the
relation takes into account only pair-correlations. Information on
higher-order correlations will necessarily reduce the entropy; thus,
the upper bound considered here will not be violated but pushed
further down. Incorporating such additional constraints into the
construction in Sec.~\ref{sec_maxH} and the derivation in
Appendix~\ref{appendix:Z} is possible.  Second, the partition function
$\cZ$ has been calculated within the simplest Gaussian
approximation. Various methods have been developed in statistical
physics to go beyond this approximation. (The exact Eq.~\eqref{Zpsi0}
is a good starting point. See also Ref.~\cite{Chakrabarty2011}.)
However, unlike the usual case where the pair-potential is given,
here, having calculated a more accurate $\cZ$, we will still need to
find the effective interaction $\lambda(\vecq)$. This will involve
integral equations, requiring numerical solution or additional
assumptions. The effects of the two approximations are hard to
decouple. We have been able to do this for the partitioned gas, whose
exact structure factor is known. Our numerical study indicates that
the error in the entropy estimation (about $22\%$) has a small
contribution from the Gaussian approximation (less than $5\%$); hence,
the main source of error in the partition-gas example lies in the
assumption of pair interactions. Improving on both approximations may
be technically difficult, but the way to go about it is known.

In view of the above, Eq.~\eqref{eq:hS} has performed remarkably
well. It is in line with the copula method for small systems and
accurately located critical points involving strong and long-ranged
correlations. The reason might lie in the variational nature of the
calculation. There are well-known examples (\eg the Self-Consistent
Expansion \cite{Schwartz2008}), where optimizing an inaccurate
expression using free parameters (here, $\lambda(\vecq)$) to match a
certain target, produces surprisingly accurate results. This is
probably why, for the partitioned gas, the relation gives reasonable
entropy estimates even though the model contains no actual
pair-interactions, and also why the Gaussian approximation introduces
such a small error there despite the strong correlations built into
the partitioned-gas model.

Another important property of the structure factor is that it
implicitly takes into account symmetries of the system such as
particle indistinguishability and translation invariance, to which the
copula splitting and data compression are blind. This significantly
reduces the relevant configuration space, thus improving the
effectiveness of random sampling.

Hyperuniformity, the strong suppression of large-scale density
fluctuations, has drawn significant attention lately
\cite{Torquato2018}. The three examples chosen for the present study
all exhibit hyperuniform states. This phenomenon is commonly defined
and characterized through the structure factor, specifically, its
decay to zero as $q\to 0$. Thus, Eq.~\eqref{eq:hS} can readily be used
to estimate the entropy cost of hyperuniformity. This may serve as an
important ingredient in future modeling of hyperuniform systems.

Our reference state throughout this work has been the ideal gas [$H=0$
  for $S(\vecq)=1$]. Certain systems may require a different reference
state. For example, a fluctuating crystal or a ``stealthy''
hyperuniform system \cite{Torquato2018}, whose $S(\vecq)$ vanishes on a
finite range of $\vecq$, have according to Eq.~\eqref{eq:hS} an
infinitely negative entropy. The natural reference in this case should
be a perfect crystal, and a different (``low-temperature'')
calculation would be needed.

This work has been focused on entropy changes related to the spatial
arrangement of particles. Materials exhibit a much wider variety of
behaviors. There are phenomena (\eg the glass transition) which leave
little to no signature on the structure factor. Entropy changes
associated with such phenomena, evidently, will not be captured by
Eq.~\eqref{eq:hS}. Nevertheless, the approach taken
here\,---\,specifying a constraint (including the proper symmetries)
and obtaining an entropy bound from an effective equilibrium
analogue\,---\,should be of more general applicability.

\begin{acknowledgments}
  We thank Roy Beck, Zohar Nussinov, and Salvatore Torquato for
  helpful suggestions. GA acknowledges support from The Israel
  Science Foundation’s Grant No.\ 373/16 and the Deutsche
  Forschungsgemeinschaft (The German Research Foundation DFG) Grant
  No. BA1222/7-1. HD acknowledges support from the Israel Science
  Foundation (Grant No.\ 986/18).
\end{acknowledgments}


\appendix

\section{Derivation of $\cZ[S]$}
\label{appendix:Z}

This appendix gives the detailed calculation leading to
Eq.~(\ref{ZGaussian}).
Our starting point is the configurational integral (\ref{ZR0}),
\begin{equation}
  \cZ = \int \left(\prod_{n=1}^N d\vecR_n\right)
  \exp \left[ -\int d\vecq \lambda(\vecq)
    \left( \sum_{m=1}^N\sum_{n=1}^N e^{i\vecq\cdot(\vecR_m-\vecR_n)} - NS(\vecq)
    \right) \right],
\label{Zsupp}
\end{equation}
and we apply a standard procedure to transform it into a functional
integral of continuous fields.

The density field is $\rho(\vecr)
= \sum_{n=1}^N\delta(\vecr-\vecR_n)$, and its Fourier transform is
$\trho(\vecq)=\sum_{n=1}^N e^{-i\vecq\cdot\vecR_n}$. We define the
conjugate field $\tpsi(\vecq)$ according to
\[
  \prod_\vecq \delta \left[\trho(\vecq) - \sum_n
    e^{-i\vecq\cdot\vecR_n} \right] = (2\pi\nu)^{-\Omega_q}
             \int\cD\tpsi \exp\left[
             i \int d\vecq \tpsi \left(\trho - \sum_n
               e^{-i\vecq\cdot\vecR_n} \right) \right].
\]
Equation (\ref{Zsupp}) is then rewritten as
\begin{equation}
  \cZ = (2\pi\nu)^{-\Omega_q} e^{N\int d\vecq \lambda S} \int
  \left(\prod_n d\vecR_n\right) \cD\trho \cD\tpsi \exp \left[ i \int
    d\vecq \tpsi \left(\trho - \sum_n e^{-i\vecq\cdot\vecR_n} \right)
    -\int d\vecq \lambda |\trho|^2 \right].
\end{equation}

The integration over the discrete configuration
$(\vecR_1,\ldots,\vecR_N)$ is taken care of through
\begin{eqnarray*}
  &&\int \left(\prod_n d\vecR_n\right) \exp \left( -i \int d\vecq
  \tpsi \sum_n e^{-i\vecq\cdot\vecR_n} \right)
  = \left[ \int
    d\vecR \exp \left( -i \int d\vecq \tpsi e^{-i\vecq\cdot\vecR}
    \right) \right]^N \\
    &&= \left\{ V + \int d\vecR \left[
    \exp \left( -i \int d\vecq \tpsi e^{-i\vecq\cdot\vecr}\right) - 1
    \right]\right\}^N \xrightarrow{N,V\rightarrow\infty,\; N/V=\brho}
  V^N \cI[\tpsi(\vecq)],
\end{eqnarray*}
where
\[
  \cI[\tpsi(\vecq)] \equiv \exp \left[ \brho \int d\vecR \left(
    e^{-i \int d\vecq \tpsi e^{-i\vecq\cdot\vecR}} - 1 \right)
    \right].
\]
We are left now with continuous fields only,
\begin{equation}
    \cZ = Z_{\rm id} (2\pi\nu)^{-\Omega_q} e^{N\int d\vecq \lambda S} \int
    \cD\trho \cD\tpsi \cI[\tpsi] \exp \left[ \int d\vecq \left( i
      \tpsi \trho - \lambda |\trho|^2 \right) \right],
\end{equation}
where $Z_{\rm id}=V^N$ (or $V^N/N!$ if the particles are
indistinguishable) is the uncorrelated (ideal gas) contribution.

We integrate over $\trho(\vecq)$ to get
\begin{eqnarray}
  \cZ &=& Z_{\rm id} (2\pi\nu)^{-\Omega_q} e^{N\int d\vecq \lambda S}
  \prod_\vecq [\pi\nu/\lambda(\vecq)]^{1/2} \int \cD\tpsi \cI[\tpsi]
  \exp\left[ -\int d\vecq \frac{|\tpsi|^2}{4\lambda} \right]
  \nonumber\\
  &=& Z_{\rm id} 2^{-\Omega_q} \pi^{-\Omega_q/2} \exp
  \left[ \int d\vecq \left( N \lambda S -
    \frac{\nu}{2} \ln(\lambda/\nu) \right) \right] Z_\psi,
  \label{Z}
\end{eqnarray}
with
\begin{equation}
  Z_\psi = \nu^{-\Omega_q} \int \cD\tpsi e^{-\cH_\psi}, \ \ \ \
  \cH_\psi = -\brho \int d\vecR \left(
    e^{-i \int d\vecq \tpsi e^{-i\vecq\cdot\vecR}} - 1 \right) +\int d\vecq 
    \frac{|\tpsi|^2}{4\lambda}.
\label{Zpsi}
\end{equation}
These expressions are exact.

We proceed with a Gaussian approximation for the integral
(\ref{Zpsi}). To order $\tpsi^2$,
\begin{equation}
  \cH_\psi \simeq i (2\pi)^d \brho \tpsi(0) + 
  \int d\vecq \frac{|\tpsi|^2}{4f(\lambda)},\ \ \ \
  f(\lambda) \equiv (\alpha^{-1} + \lambda^{-1})^{-1},
\label{Hpsi1}
\end{equation}
where $\alpha\equiv [2(2\pi)^d\brho]^{-1}$.  Omitting the $\vecq=0$
term and integrating over $\tpsi$, we get
\begin{equation}
  Z_\psi \simeq \prod_\vecq \left[ (4\pi/\nu)f(\lambda) \right]^{1/2}
  = (4\pi)^{\Omega_q/2} \exp \left(
  \frac{\nu}{2} \int d\vecq \ln \left[f(\lambda)/\nu \right] \right).
\label{Zpsi1}
\end{equation}
Substitution in Eq.~(\ref{Z}) gives
\begin{equation}
  \cZ = Z_{\rm id} \exp \left[ \int d\vecq \left( N\lambda S +
    \frac{\nu}{2} \ln \left[f(\lambda)/\lambda \right] \right) \right],
\label{Z1}
\end{equation}
which coincides with Eq.~(\ref{ZGaussian}).

\section{Implementation of statistical entropy estimations}
\label{appendix:entropy_estimation}

\subsection{Direct statistical estimation}

As discussed in section~\ref{sec_copula}, we apply a recently
suggested method that is based on decomposing the distribution into a
product of the marginal distributions and the joint dependency, also
known as the copula \cite{Ariel2020}.  Recall Sklar's theorem
\cite{Book:copula,Durante2010}, which states that any continuous
multi-dimensional density $p(\vecx)$, $\vecx \in \R^D$ can be written
uniquely as
\begin{equation}
       p(\vecx) = p_1(x_1) \cdot \dots \cdot p_D (x_D) c(F_1(x_1),\dots,F_D(x_D)) ,
\end{equation}
where, $\vecx=(x_1,\dots,x_D)$, $p_k(\cdot)$ denotes the marginal
density of the $k$'th dimension with cumulative distribution function
(CDF), $F_k (t) = \int_{-\infty}^t p_k(x) dx$, and $c(u_1,\dots,u_D)$
is the density of the copula, i.e., a probability density on the
hyper-square $[0,1]^D$ whose marginals are all uniform on $[0,1]$.

Analytically, the copula is obtained by a change of variables,
$u_k=F_k(x_k)$.  This motivates estimating the copula as follows.  Let
$\vecx^i = (x^i_1, \dots , x^i_D) \in \R^D$, where $i=1 \dots N$
denote $N$ independent samples from a real $D$-dimensional random
variable with density $p(\vecx)$.  We would like to use the samples
$\vecx^i$ in order to obtain samples from the copula density $c(u_1,
\dots, u_D)$.  This can be obtained by finding the rank (in increasing
order) of samples along each dimension.  Dividing by $N$, we obtain,
$\vecu^i = (u^i_1, \dots , u^i_D) \in [0,1]^D$, were $u^i_k = n/(N+1)$
if $x^i_k$ is the $n$ smallest value out of the list $u^1_k, \dots ,
u^N_k$.  Overall, we obtain $N$ samples from the distribution
$c(u_1,\dots,u_D)$.

\begin{figure}[h!tp]
\centering
\includegraphics[trim={3cm 5.5cm 3cm 0cm},clip, width=9cm, angle=0]{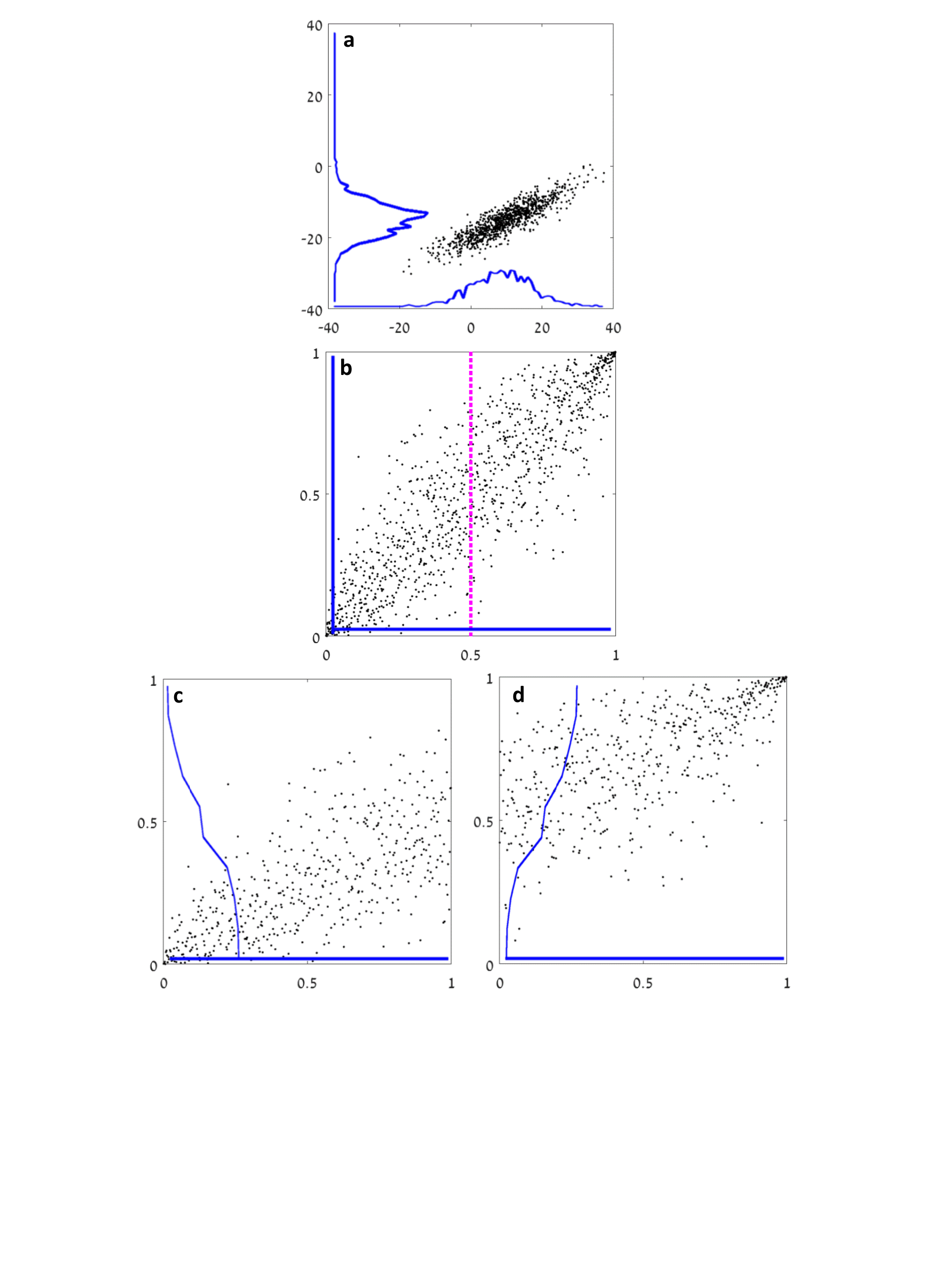}
\caption{ A schematic sketch of the copula splitting method,
  reproduced from Ref.~\cite{Ariel2020}.  {\bf a.} A sample of 1000 points
  from a 2D Gaussian distribution. The blue lines depict the empirical
  density (obtained using uniform bins).  {\bf b.} Enumerating the
  sorted data in each dimension (and dividing by $N$), the same data
  provides samples for the copula in $[0,1]^2$.  Splitting the data
  according to the median in one of the axes (always at 0.5) yields
  {\bf c} (left half) and {\bf d} (right half).  The blue lines depict
  the empirical marginal densities in each half.  Continue
  recursively.  }
\label{fig:copulasketch}
\end{figure}

In order to approximate the entropy, 1D entropy of marginals is
estimated using either uniform binning or sample-spacing methods,
depending on whether the support of the marginal is known to be
compact (bins) or unbounded/unknown (spacing).  The main challenge
lies in evaluating the differential entropy of high-dimensional
copulas \cite{Calsaverini2009,Embrechts2013}.  Here, we compute it
recursively, similar to the $k$DP approach \cite{Stowell2009}.  Let $k
\in \{1,\dots,D \}$ be one of the
dimensions, to be chosen using any given order.  The copula samples
$u^i$ are split into two equal parts (note that the median in each
dimension is always $1/2$).  Denote the two halves as $v_j^i = \{
u_j^i | u_k^i \le 1/2 \}$ and $w_j^i = \{ u_j^i | u_k^i > 1/2 \}$.
Scaling the halves as $2v_j^i$ and $2w_j^i-1$ produces two sample sets
for two new copulas, each with $N/2$ points.  A simple calculation
shows that
\begin{equation}
       H_c = \frac12 ( H_{2v} + H_{2w-1} ) ,
\end{equation}
where $H_{2v}$ is the entropy estimate obtained using the set of
points $2v_j^i$ and $H_{2w-1}$ is the entropy estimate obtained using
the set of points $2w_j^i-1$.  The marginals of each half may no
longer be uniformly distributed in $[0,1]$, which suggests continuing
recursively, i.e., the entropy of each half is a decomposed using
Sklar's theorem, etc.  See Fig.~\ref{fig:copulasketch} for a schematic sketch
of the method.  See Ref.~\cite{Ariel2020} for convergence analysis and
numerical examples.

A Matlab code is available in MathWorks MATLAB Central.

%


\subsection{Compression-based estimation}

As discussed in Sec.~\ref{sec_zip}, we estimate the entropy using a
lossless compression algorithm.  Similar to
Ref.~\cite{AvineryPRL2019}, samples are binned into 256 equal bins in
each dimension, and the data are converted into a matrix of 8-bit
unsigned integers, in which each row constitutes an independent sample.
The matrix is compressed using the LZW algorithm (implemented in
Matlab's imwrite function) to a gif file.

Specifically, suppose we are given $K$ samples, $x_1,\dots,x_K$, in a
$D$-dimensional box $[0,L]^D$.  The algorithm we apply is as
follows.
\begin{enumerate}
   \item Digitize the samples into 256 equally sized bins, $y_k=\llcorner 255x_k/L\lrcorner$, where $\llcorner \cdot \lrcorner$ denotes rounding down to the nearest integer. Construct the $K \times D$ image matrix $Y$ which has $y_1,\dots,y_K$ as rows.
   \item Convert the image $Y$ into a gif file. Denote the file size $C_{\rm data}$.
   \item Convert an image $Y_0$, a $K \times D$ matrix of zeros (8-bit unsigned integers) into a gif file. Denote the file size $C_0$.
   \item Convert an image $Y_1$, a $K \times D$ matrix of random integers between 0 and 255 (8-bit unsigned integers) into a gif file. Denote the file size $C_1$.
   \item Interpolation: Let $\lambda=(C_{\rm data}-C_0)/(C_1-C_0)$ and
     take $H=\lambda H_0 + (1-\lambda) H_{\rm ideal}$.  Here,
     $H_0=D(\ln L - \ln 255)$, the maximum entropy of a random
     variable that yields a sample matrix $Y_0$ (independent uniform
     distribution in $[0,L/255]$), and $H_{\rm ideal}=-D \ln L$ is the
     entropy of an ideal gas of $D$ particles in $[0,L]$.
\end{enumerate}

\section{Numerical optimization of $\lambda(q)$}
\label{appendix:max_entropy_model}

This appendix gives further details concerning the optimization
results presented in Sec.~\ref{sec_gas}.  A similar method was
presented in \cite{Zhang2020}.  To optimize $\lambda(q)$ we need to
solve the system of equations resulting from the Lagrange multipliers,
Eqs.~\eqref{eq:appA}. The main idea is that, given values
$\lambda(\vecq_1),\dots,\lambda(\vecq_M)$ for a subset of wavevectors
$\vecq$, one can compute the structure factor (assuming all other
$\lambda$'s are zero).  Since the structure factor of a distribution
of the form of Eq.~\eqref{eq:appA:pj} is unique \cite{Henderson1974},
we numerically find the set of $\lambda$'s that minimize the distance
from $S(q)$ given by \eqref{eq:appA:Sq_supp} and the exactly known
$S_{\rm PG}(q)$. Then, the entropy is approximated by numerically
computing $\cZ$ of Eq.~\eqref{eq:cZdisc}.  Thus, the numerical
optimization method finds the entropy of the maximum-entropy model,
eliminating the error incurred by the second-order expansion in
Sec.~\ref{sec_eff} and Appendix~\ref{appendix:Z}. This numerical
approach is only applicable for low-dimensional systems.

We write \eqref{eq:appA:pj}, the probability of a configuration $s$,
as
\begin{equation}
  p_s = \hat{\cZ}^{-1} \exp \left[ -  \sum_{\vecq \neq 0} \lambda(\vecq) \sum_{m,n} e^{i\vecq\cdot(\vecR_m^s-\vecR_n^s)}  \right] ,
\label{eq:appC:pj}
\end{equation}
where we absorbed the exponential term including the structure factor into the normalization constant,
\begin{equation}
   \hat{\cZ} = \cZ \exp \left[ - N \sum_{\vecq \neq 0} \lambda(\vecq)  S(\vecq) \right] .
\end{equation}
The sum (or integral) over configurations $s$ is approximated using a
naive Monte-Carlo method as follows.  First, approximate the infinite
sum over $\vecq$ by a finite number, $\vecq = (q_0=0, q_1,
2q_1, \dots \Omega_q q_1)$, where $q_1=2 \pi/N$ is the smallest
(non-zero) wavenumber.  We generate $K$ independent random
configurations from an ideal-gas distribution.  In the
partitioned-gas case with $N$ particles, $R_n^s$ are independent,
uniformly distributed in $[0,N]$, $s=1\dots K$, $n=1 \dots N$.  Denote
\begin{equation}
  \alpha_s =  \sum_{0 \neq \vecq \in \Omega_q} \lambda(\vecq) \sum_{m,n} e^{i\vecq\cdot(\vecR_m^s-\vecR_n^s)},\ \ \ \ 
  \tilde\alpha_s = \alpha_s - \langle \alpha_s \rangle,\ \ \ \ 
 \tilde{p}_s =  \tilde\alpha_s  / \sum_s  \tilde\alpha_s .
\end{equation}
%
%
Thus, $\tilde{p}_s$ is a numerical approximation of $p_s$.  Note that
we have centered the exponential factors $\alpha_s$ to avoid
excessively large or small exponents.  In the above,
$\langle \cdot \rangle$ denotes averaging over all samples $s$.
Substituting into the structure factor, Eq.~\eqref{eq:appA:Sq_supp},
while replacing $p_s$ with $\tilde{p}_s$, we obtain an approximation
for the structure factor $S(\vecq)$.  Thus, given a set of Lagrange
multipliers $\lambda(\vecq)$, $q \in \Omega_q$, we can compute the
resulting structure factor $\tilde{S}(\vecq)$, assuming that
$\lambda(\vecq)=0$ for all $\vecq \notin \Omega_q$.

Our goal is to find a set of $\lambda(\vecq)$, $q \in \Omega_q$, such
that $\tilde{S}(\vecq)$ is close to the required $S(\vecq)$.  Here, we
applied Matlab's non-linear least square fitting function lsqcurvefit
using a zero initial guess.  Figure~\ref{fig:bestLambda} shows results
with $N=10$. Parameters are $\Omega_q=10N$ and $K=1000N \Omega_q$.  
The method only works for relatively small $N$
(up to about 10).  At higher dimensions, the naive Monte-Carlo
sampling fails and more sophisticated methods biasing the sampling
toward high-probability regions are necessary.

Next, we approximate the entropy $H$.  Here, care must be taken when
converting integral to sums, as prefactors change the entropy.  Recall
the partition function, Eq.~\eqref{ZR0} or \eqref{Zsupp}.
%
%
Several approximations are required: (a) Integration over
configurations is replaced by Monte-Carlo sampling.  (b) The integral
over $\vecq$ is replaced with a sum over a finite number $\Omega_q$ of
wavenumbers.  (c) The structure factor $S(\vecq)$ is replaced with its
approximation $\tilde{S}(\vecq)$.
Overall, $\cZ$ is approximated as
\begin{equation}
  \tilde\cZ = K^{-1} \sum_{s=1}^K
  \exp \left[ - \Delta q \sum_{0 \neq q \in \Omega_q} \lambda(\vecq)
    \left( \sum_{m=1}^N\sum_{n=1}^N e^{i\vecq\cdot(\vecR_m-\vecR_n)} - N \tilde{S} (\vecq)
    \right) \right],
\end{equation}
where $\Delta q = 2 \pi/N$.
Finally, the entropy $H = \ln \tilde\cZ$ is computed as.
\begin{eqnarray}
&& \beta_s  =  \Delta q \sum_{0 \neq q \in \Omega_q} \lambda(\vecq)
    \left( \sum_{m=1}^N\sum_{n=1}^N e^{i\vecq\cdot(\vecR_m-\vecR_n)} - N \tilde{S} (\vecq)
    \right),
    \nonumber \\
    && \tilde\beta_s = \beta_s - \langle \beta_s \rangle,
    \nonumber \\
    && H = \ln \tilde\cZ = \ln \left( \sum_s e^{-\tilde\beta_s} \right) - \ln K - \langle \beta_s \rangle.
\end{eqnarray}
Once again, to reduce round-off errors, arguments are centered before
taking the exponent.


\bibliography{entropy_refs}

\end{document}